# A technology agnostic RRAM characterisation methodology protocol

Spyros Stathopoulos, Loukas Michalas, Ali Khiat, Alexantrou Serb and Themis Prodromakis

[1]Electronic Materials & Devices Research Group, Zepler Institute
University of Southampton, SO17 1BJ, Southampton, UK

*Corresponding Author: Dr Spyros Stathopoulos (Email: s.stathopoulos@soton.ac.uk)

**The emergence of memristor technologies brings new prospects for modern electronics via enabling novel in-memory computing solutions and affordable and scalable reconfigurable hardware implementations. Several competing memristor technologies have been presented with each bearing distinct performance metrics across multi-bit memory capacity, low-power operation, endurance, retention and stability. Application needs however are constantly driving the push towards higher performance, which necessitates the introduction of standard characterisation protocols for fair benchmarking. At the same time, opportunities for innovation are missed by focusing on excessively narrow performance aspects. To that end our work presents a complete, technology agnostic, characterisation methodology based on established techniques that are adapted to memristors/RRAM characterisation needs. Our approach is designed to extract information on all aspects of device behaviour, ranging from deciphering underlying physical mechanisms to benchmarking across a variety of electrical performance metrics that can in turn support the generation of device models.**

Resistive switching devices, also known as memristors[1], have exhibited an unmatched potential for a broad range of applications ranging from non-volatile memories[2] to neuromorphic computing[3,4] and reconfigurable circuits[5,6]. As the scope of the resistive memories expands there is the need for a holistic testing and electrical characterisation methodology that covers all aspects of the performance evaluation of the device under test. In this paper we propose a comprehensive characterisation protocol for resistive memory cells that is technology agnostic, i.e. supports equally well volatile/non-volatile and binary/analogue memory technologies. Our approach also allows capturing signatures that are indirectly related to their performance, such as the physical mechanisms enabling resistive switching. Starting with the pristine device we link the I-V characteristics with the underlying conduction mechanisms followed by the electroforming process. As the electroforming fundamentally affects the material properties of the stack the conduction mechanisms need to be re-evaluated. Next part of the protocol are the basic endurance/retention tests. Finally we assess the memory capacity of the cell followed by a phenomenological model extract. All the measurements have been performed with our in-house developed memristor characterisation platform, ArC ONE[7].

## I-V characterisation of pristine devices

Memristors are two-terminal Metal-Oxide-Metal (MOX) stacks that typically obtain their memristive nature after the electroforming process. However analysis of the I–Vs in their pre-formed, or the so called pristine, state could be a very useful initial characterisation step. Recording of the pristine state I–V provides i) a *straightforward testing* for the device operation ii) indications for any type (volatile or stable) of *forming-free switching ability* and iii) a deeper insight on the *device physics*, i.e. the mechanism underneath the electrical response. Potential straightforward correlation between the pre- and post-formed characteristics, not yet achieved, would be extremely useful as it will enable performance optimization through design and fabrication.

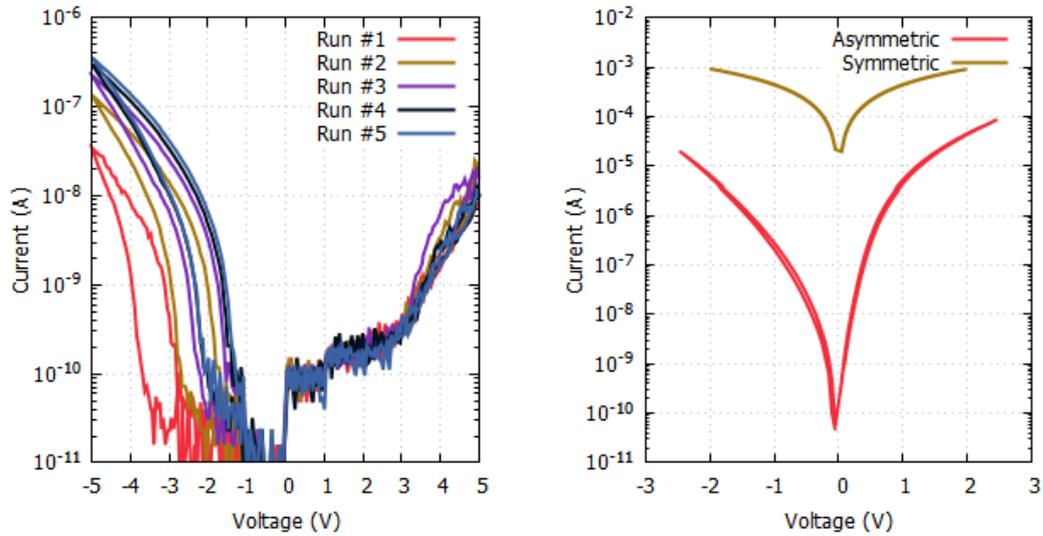

**Fig. 1**: Pre-forming current – voltage characteristics. (left) Hysteresis loop and drift of the I–V arise by the multiple acquisition iterations. This behaviour denotes strong ionic character in the core metal oxide thin film. (right) Typical asymmetric (red) and symmetric (yellow) I–Vs of a Au/TiO$_2$/Pt device and a Au/TiO$_2$/Al respectively, obtained for interface and material controlled transport respectively in Metal-Oxide-Metal stacks.

Regarding the responsible mechanisms, the I–V is the most valuable tool to reveal the underlying physics. An unstable/non reproducible I–V (fig. 1a), could be associated with important contribution of movable ions, associated to the oxygen vacancies spontaneously generated in the transition metal oxides. Movable ions could also results in stable I–Vs presenting notable hysteresis loops on a full acquisition cycle. The ions contribution to the device conduction could be direct or indirect, for example by modifying the charge close to the interface, thus the potential barrier accordingly, resulting in a conductivity change.

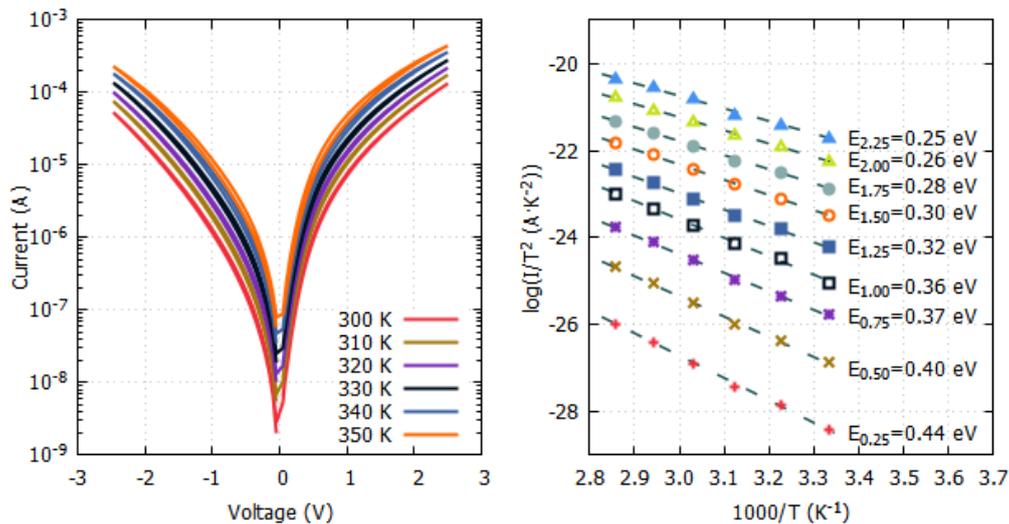

**Fig. 2**: Temperature analysis for pristine Metal-Oxide-Metal stacks depicting evolution of I–V characteristics with increasing temperature (left) and temperature dependent signature plots (right) deriving from them. This analysis allows for determining the dominant transport mechanism. This particular case corresponds to thermionic emission over the interface-formed Schottky barrier[8].

Stable and reproducible I–Vs can be further assessed. An asymmetric response with respect to the bias polarity is a strong indication of transport determined by the interface barriers whilst a symmetric curve should be associated with core-material controlled transport (fig. 1b).

Despite this, the in depth clarification of the transport properties requires assessing the temperature dependence of the I–Vs. This is because the various conduction mechanisms obey both field and temperature dependence[9] and thus the recording of the I–V curves at different temperatures (fig. 2a) will allow the extraction of characteristic signature plots. This could be simple Arrhenius plots or more sophisticated when the dominant mechanism involves processes such as hopping, Frenkel-Poole emission[10] or thermionic emission over an interface Schottky barrier[8]. Further assessment of the signature plots allows for extraction of quantitative results for parameters such the interface barrier height or the activation energy of the involved defects in the cases of interface and material controlled transport respectively.

## The electroforming process

In order for the devices to be practically exploitable as tunable resistors (aka memristors), an electroforming step is usually required. Electroforming was originally mentioned by Hickmott[11,12] where he describes it as an irreversible change of the electrical properties of the material by applying a voltage greater than a minimum *forming* voltage. Depending on the RRAM technology used the electroforming process can represent either the formation of a conductive filament due to structure altering effects or oxygen deprivation[13], the diffusion of metal into the metal oxide layer or the lowering of the interfacial barrier between the electrodes and the metal oxide active layer[14].

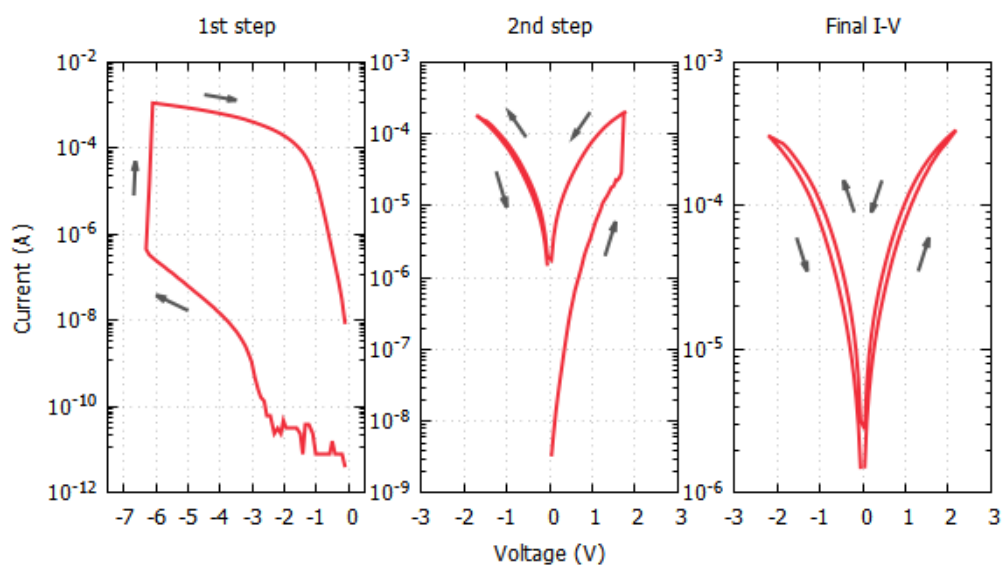

**Fig 3**: Two step electroforming using staircase I-V of a Pt/TiO$_2$/Pt device. Apparent electroforming voltage from the GΩ range is ~-6.5 V (left). The device is undergoing a further electroforming step bringing it down to the ~30 kΩ range (middle) where a steady state is established (right).

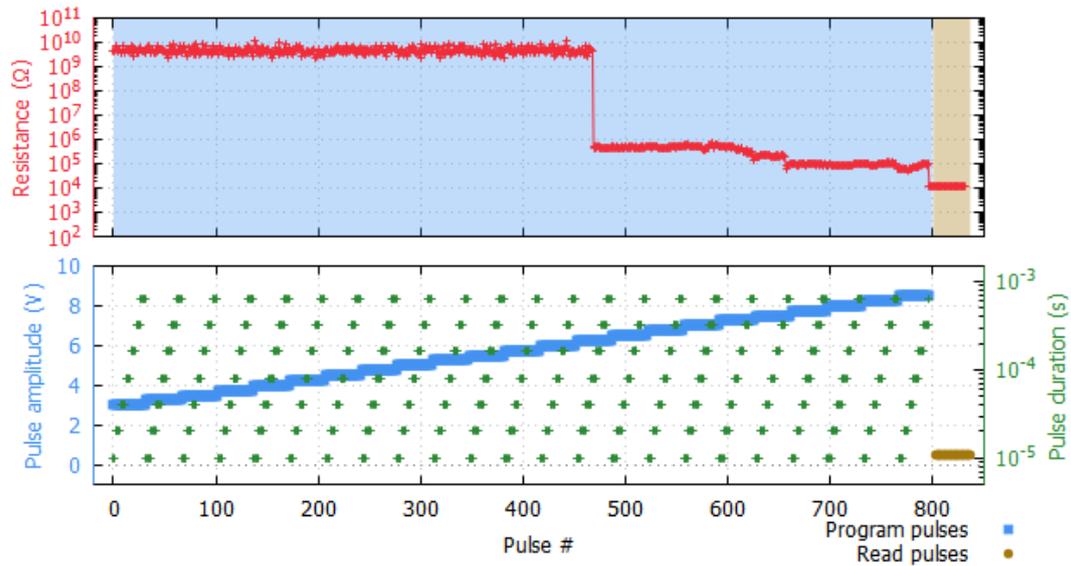

**Fig 4**: Typical two step electroforming process using a pulse sequence. This device is formed with pulses of increasing amplitude and pulse width. Target resistance threshold has been set to 10 kΩ. The device initially drops to the MΩ range before attaining its final value. Highlighted in blue is the biasing region, followed by a series of READ pulses (highlighted in yellow).

Regardless of the actual forming mechanism during the process the resistance of the device is usually lowered to ranges that are typically relevant to applications. It is apparent that electroforming is not possible before a device-dependent voltage threshold is reached. However the way to cross that threshold is not immediately obvious. One way to form a device is during an I–V cycle as can be seen in fig. 3a. The device is biased with increasingly higher voltage steps until an immediate change in the resistive state is observed. Further steps might be required to move the device to non-volatile region. As can be seen from fig. 3b the conductance of the device is still lower than the one achieved in the first step. So a further forming cycle is required (fig. 3b) before reaching a final state (fig. 3c). As further increasing the voltage amplitude leads to partial dielectric breakdown[15] of the active layer *compliance* is an issue that must be dealt with during electroforming in order to prevent irreversible switching degeneration of the device. Two distinct forms of compliance can be identified: *current compliance* and *time compliance* (ie. short controllable pulses). The key issue with the former mode is that there is a distinct delay before reaching the current cut-off threshold with current overshoot, partly due to delayed compliance system reaction and partly due to residual parasitic capacitance[16], posing a significant problem. As such, short sequential pulses can be a better alternative for a controllable forming procedure. Instead of continuously biasing the device, sequential voltage pulses of continuously increasing duration and amplitude are applied to the device. The result is a more uniform procedure of attaining the desired formed state. In the example shown in fig. 4 a series of programming pulses from 3 to 10 V is applied to the device. Within each step the pulse duration is modulated from the low-μs up to the ms range.

In addition fig. 4 further illustrates the common pattern in the electroforming process where there is an initial increase in the electrical conductivity of the device which marks the onset of the electroforming mechanism. By continuing the application of the programming pulses the device is finally driven to its target resistive state. While after the first forming the device is able to be tuned to a multitude of resistive states the initial increase to the conductance of the device is irreversible as it is associate with morphological alterations of the active layer[17]. This behaviour is also evident in the electroforming procedure using staircase I-Vs (fig. 3).

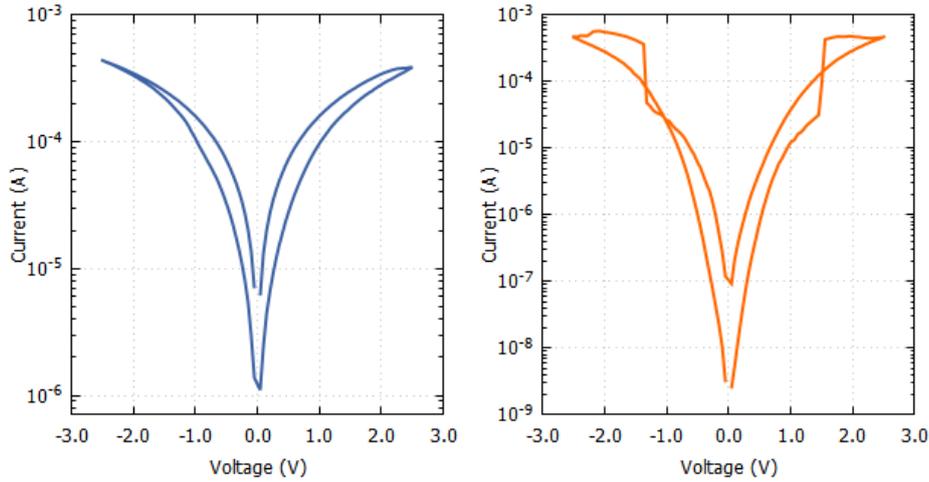

**Fig. 5**: Post-forming I–V characteristics exhibiting typical bipolar behaviour (left; opposite bias polarity is required to SET/RESET the device resistive states) and unipolar behaviour (right; the device SET/RESET by increasing the applied bias even at the same polarity).

## Post electroforming I-V characterisation

After the electroforming process the device typically obtain its memristive nature. At this point the core material and the interfaces of the stack have been completely reformed with respect to their pristine state. It is therefore essential to commerce the post forming characterisation by acquiring once more the non-invasive current vs voltage characteristics. This analysis will provide all the information discussed in the previous part, such as the stability/reproducibility of the device characteristics (many post forming conditions are not stable – referred to as

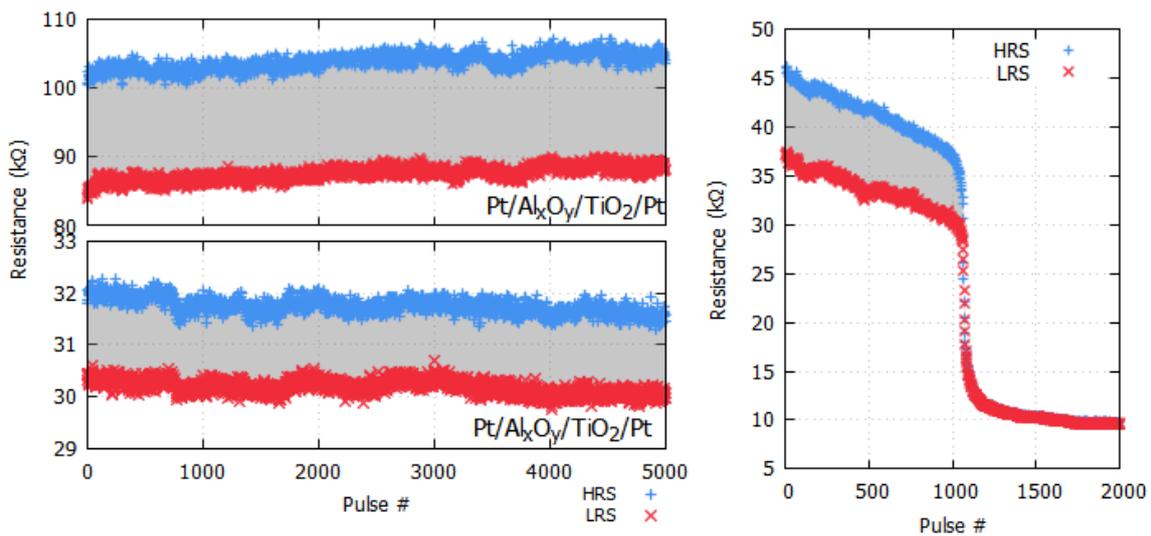

**Fig. 6**: Endurance characterisation of two different devices. On the left a Pt (TE)/Al$_x$O$_y$/TiO$_2$/Pt (BE) is subjected to 5000 pulses of 2 V pulses of alternating polarity for two different resistive ranges. In both cases the resistive window is maintained throughout. On the right figure a device failing after 1000 pulses.

*volatile*), the operation limits and physical mechanism underneath the conduction process, by involving characterisation at different temperatures. Further to this, recording of the I-Vs will provide an initial indication of the resistive switching character i.e. bipolar (fig. 5a) or unipolar (fig. 5b), whilst it leads to detail recording of the SET/RESET voltages. These indications can be more accurately recorded by more specialised testing (see section "Switching dynamics"). This is important information to be used also on the following characterisation steps, such as the endurance and/or multi state seeking, described in more details in the following sections.

## Base Performance evaluation (Endurance/Retention)

A basic characteristic of a RRAM cell is its ability to alternate between two resistive states. In order to determine how the devices behave under repeated stress an endurance test is required. For a typical bipolar device a series of alternating polarity pulses is applied to the device switching it between neighbouring resistive states.

A typical output of a device under this test can be seen in fig. 6 where 2 V 100 μs pulses of alternating polarity are applied to a Pt/Al$_x$O$_y$/TiO$_2$/Pt device for two different resistive ranges. A well behaving device retains its memory window while subject to such biasing. A failing device (also evident in fig. 6) will have its memory window, created between a low (LRS) and high resistive state (HRS), quickly deteriorating to a point that the memory window is completely eliminated.

Although the endurance test exhibits the ability of the device to switch reliably between two states it does not account for the longevity of the two states once they are established. In order to do so a retention step is required. In fig. 7 the device is continuously read for a period of up to three hours for the two different resistive ranges as defined in fig. 6. The data are then extrapolated for a period of up to 10 years (~10$^6$ minutes) so that the stability of the memory window can be established. Retention testing can be performed at any state within the operating range of the device.

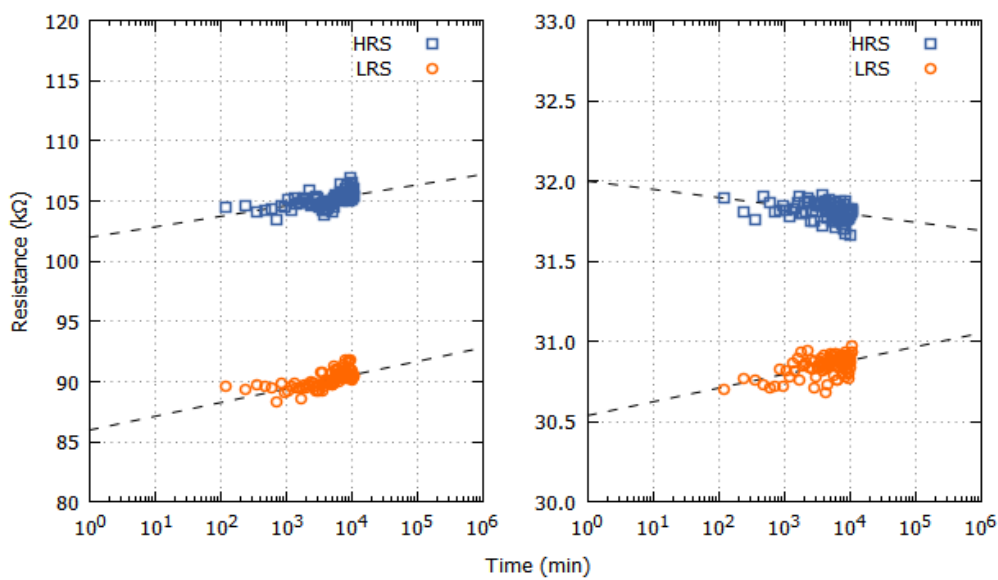

**Fig. 7**: Retention measurements for the two different resistive states defined in fig. 6. Read voltage set at 0.2 V. By extrapolating the retention sampling data the device on the left retains its memory window despite the drift in resistance, whereas the device on the right has seen a deterioration to its window for the same period.

## Switching dynamics

Important information can be gathered from the device to evaluate its switching dynamics, ie. the switching range, the switching polarity as well as its behaviour given a specific stimulus which allows for further optimisation of the biasing scheme applied. To do so a two-stage characterisation algorithm can be used[18]. Initially pulses of alternating polarity are applied to the device in order to determine the direction of change in the conductance of the device given the stimulus and then voltage ramps are used to determine the actual change in the conductance.

A typical output of such routine can be seen in fig. 8 where a series of 100 μs pulses of increasing amplitude has been applied to two different Pt/TiO$_2$/Pt and Au/TiO$_2$/Pt devices operating at the ~30 kΩ and ~5 kΩ range respectively. Given these results it is possible to establish a switching regime for the two devices. The first has an evidently bipolar response to pulses of different polarity which is the typical behaviour expected for a bistable memory device. However, the second device exhibits a hybrid bipolar/unipolar behaviour where the applied bias can result to both an increase and a decrease of the device conductance depending on the amplitude when the applied bias surpasses a threshold and the behaviour of the device changes to unipolar. Although in the first case the operating boundaries of the device are clear in the second case this analysis, and given the application at hand, allows us to establish an operating voltage range where the device remains strictly within the bipolar switching regime.

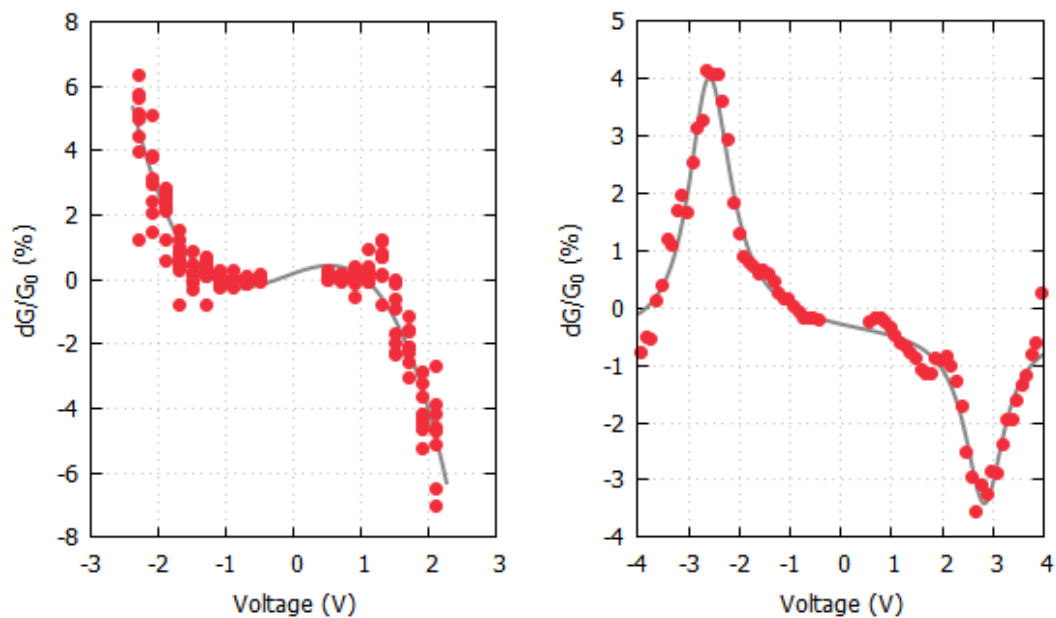

**Fig. 8**: Relative conductance change in response to a given stimulus. Typical response of a Pt/TiO$_2$/Pt device exhibiting bipolar behaviour (left) and a Au/TiO$_2$/Pt exhibiting a hybrid unipolar/bipolar behaviour (right).

## From devices to applications – Memory capacity

All of the previous steps mostly focus on the evaluation of the performance of the device itself. Broaching into the area of applications of particular importance is the ability to identify the memory capacity of the device. Although the bistable aspect of the device operation is straightforward give the capability of the memristor to variate between two extreme states exposing further resistive states warrants a more targeted characterisation sequence. The need for multibit memory capacity in conventional RRAM cells is of utmost importance to a series of applications ranging from non-volatile memories to reconfigurable circuits and neuromorphic computing.

Towards that end a comprehensive characterisation routine can be used as described in a previous publication[19]. A succession of fixed pulse width pulse trains each containing an increasing number of programming pulses is applied to the device followed by a short retention test to assess the stability of the current resistance. If during this test the resistance of the device remains within a tolerance band a new state is registered. If not the amplitude of the bias is increased up to a specified limit and a new succession of pulse trains in applied. In figure 9 we can see a typical multibit characterisation output for a Pt/TiO$_2$/Pt RRAM cell. Read voltage is kept at 0.5 V and the characterisation protocol is applied for up to 10 pulses of 1 µs programming pulses ranging from 1.6 V to 2.1 V with a confidence interval of 2σ. In this example, 5 bits of information (31 states) can be extracted from the device within just 4 kΩ of resistance span (8 to 12 kΩ). Depending on the target application the confidence intervals can be adjusted to allow for either a high number of states or a larger interval in terms of resistance between the states. As such the effective range of a device can be tiled in an efficient way to accommodate for the demands of the application at hand.

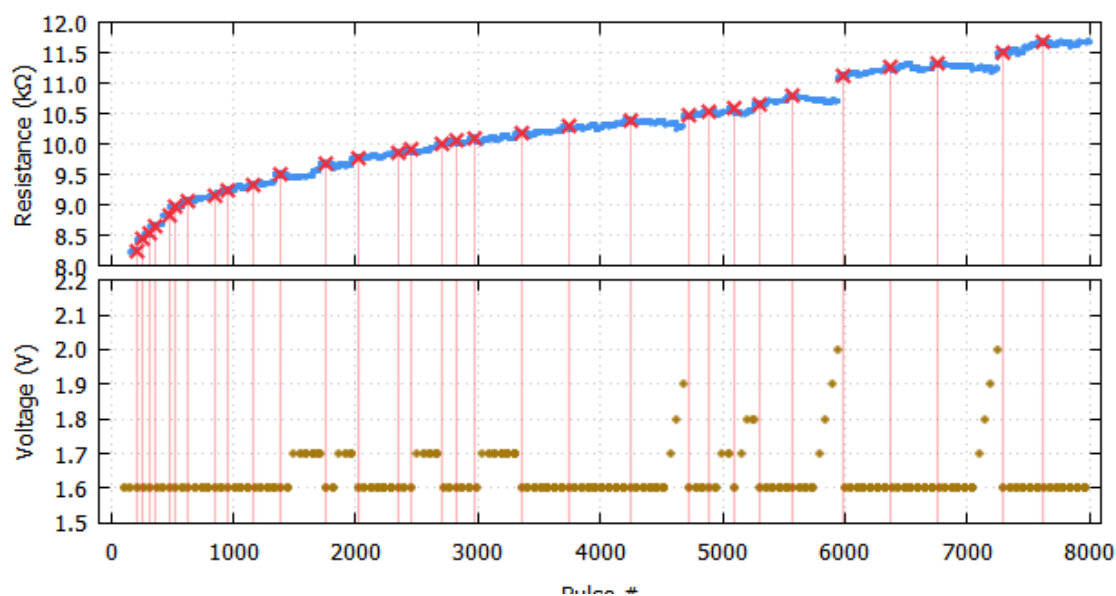

**Fig. 9**: Multibit evaluation of a Pt/TiO$_2$/Pt RRAM cell. In the top figure stable resistive states have been marked with red crosses. The corresponding programming protocol is shown in the bottom figure. Read pulses are applied continuously throughout and are set at 0.5 V.

| Positive pulses | Negative pulses | Units |
|---|---|---|
| $A_p = -0.14$ | $A_n = 0.02$ | $\Omega \cdot s^{-1}$ |
| $t_p = 2.74$ | $t_n = 3.03$ | V |
| $a_{0p} = 24.4 \times 10^3$ | $a_{0n} = 14.7 \times 10^3$ | $\Omega$ |
| $a_{1p} = -4.75 \times 10^3$ | $a_{1n} = -2.33 \times 10^3$ | $\Omega \cdot V^{-1}$ |

Table 1: Fitting parameters for the phenomenological model used in this example.

## From devices to applications – Phenomenological modelling

In order to properly integrate RRAM into an integrated circuit workflow it is necessary to have realistic, accurate and computationally efficient behavioural models extracted from readily available data. By analysing the transient response of the device under continuous bias using pulse trains of varying amplitudes and alternating polarities it is possible to extract an accurate phenomenological model for a device where the rate of change of the resistance of the device is modelled a function of the resistance itself ($R$) and the applied bias ($v$)[20].

$$\frac{dR}{dt} = s(v) \times f(R, v)$$

where $s(v)$ is the switching sensitivity and $f(R, v)$ the window function.

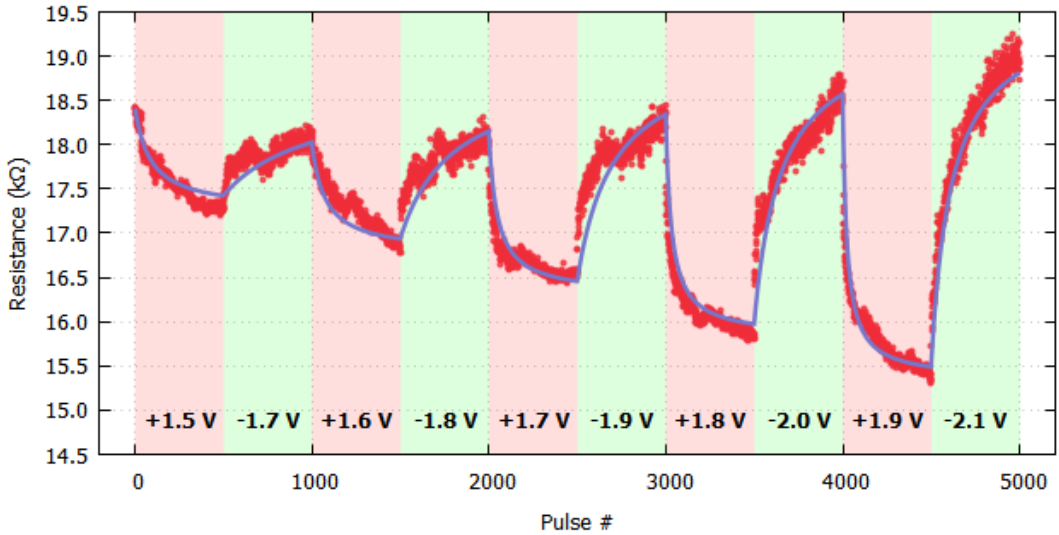

Fig 10: Analytical model (solid blue line) extracted from the resistive response of a Pt/TiO2/Pt RRAM cell (red points) using 500 pulse batches of alternating polarities. The amplitude of each of the pulse trains applied is indicated on the bottom of the graph. The initial resistance of the device is 18.3 kΩ.

$$s(v) = \begin{cases} A_p\left(-1 + \exp\left(\frac{|v|}{t_p}\right)\right), v > 0 \\ A_n\left(-1 + \exp\left(\frac{|v|}{t_n}\right)\right), v < 0 \end{cases} \text{ and } f(R,v) \begin{cases} (a_{0,p} + a_{1,p}v - R)^2, v > 0 \\ (R - a_{0,n} + a_{1,n}v)^2, v < 0 \end{cases}$$

The rest of the parameters are free fitting variables for the positive and negative branch of the bias. The model can be readily applied to data obtained during the characterisation procedure by applying a fixed number of programming pulses while alternating the polarities and fitting the above equations in a least square fashion. An example of such application can be observed in fig. 9 where a bipolar Pt/TiO2/Pt RRAM cell is biased with alternating programming pulses of increasing amplitude ranging from 1.5 V to 1.9 V and -1.7 V to -2.1 V. The fitting parameters for this device are summarised in table 1.

## Conclusions

In this paper we presented a complete testing methodology to address the characterisation of memristive cells. Our routine covers all technological facets required to integrate a new resistive memory technology on any workflow, starting from fundamental physical aspects to performance characteristics memory capacity and behavioural modelling. Figure 11 summarises the aspects that were discussed in this article.

## Acknowledgements

The authors would like to acknowledge financial support from Engineering and Physical Sciences Engineering Research Council (EPSRC) programme grants no. EP/K017829/1 (Reliably unreliable nanotechnologies) and EP/R024642/1 (Functional Oxide Reconfigurable Technologies – FORTE).

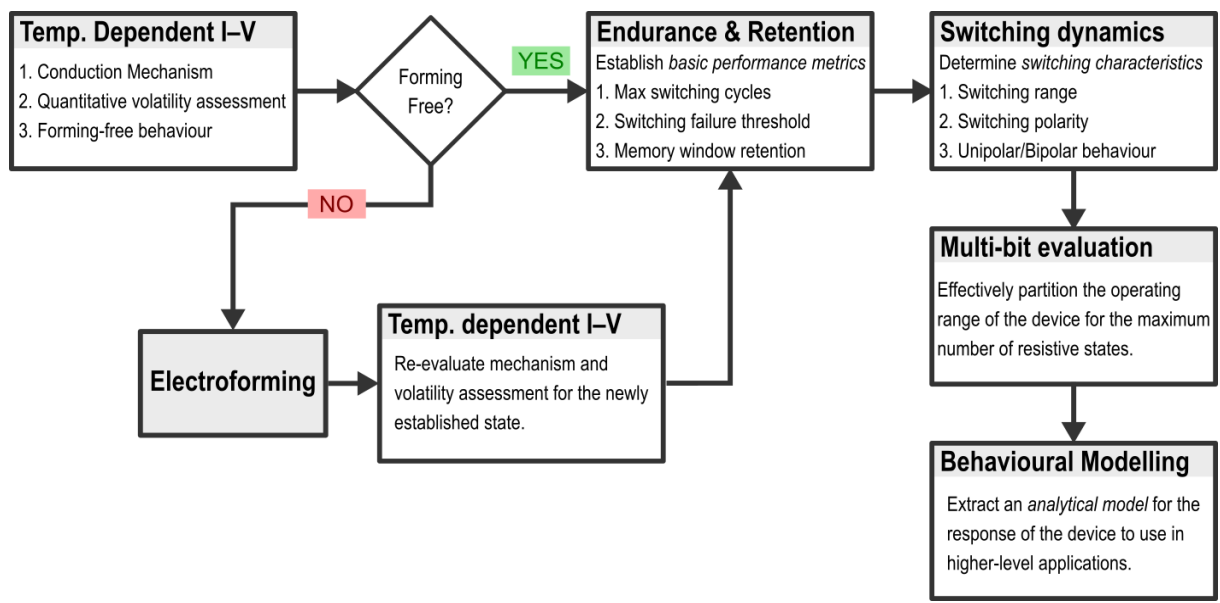

**Figure 11**: The complete proposed characterisation routine as discussed in this article.